\documentclass{esp}
\usepackage{times}
\usepackage{graphicx}
\usepackage{hyperref}
\usepackage{url}
\usepackage{booktabs}
\usepackage[normalem]{ulem}
\usepackage{xspace}
\usepackage{multicol}
\usepackage{float}
\usepackage{natbib}
\usepackage{titletoc}

\usepackage{amsmath,amsfonts,bm}

\def\eqref#1{equation~\ref{#1}}

\def\1{\bm{1}}

\def\va{{\bm{a}}}

\def\vh{{\bm{h}}}

\def\vq{{\bm{q}}}

\def\vx{{\bm{x}}}
\def\vy{{\bm{y}}}
\def\vz{{\bm{z}}}

\DeclareMathAlphabet{\mathsfit}{\encodingdefault}{\sfdefault}{m}{sl}
\SetMathAlphabet{\mathsfit}{bold}{\encodingdefault}{\sfdefault}{bx}{n}

\newcommand{\cmmnt}[1]{}
\newcommand\method{NatureLM-audio\xspace}
\newcommand\newbeans{BEANS-Zero\xspace}

\useunder{\uline}{\ul}{}

\title{\method: an Audio–Language \\ Foundation Model for Bioacoustics}

\title{\method:\\ an Audio–Language  Foundation Model for Bioacoustics}
\shorttitle{\method}
\author{%
  David Robinson$^{\star}$}
  \author{
  Marius Miron$^\dagger$}
  \author{Masato Hagiwara$^\dagger$}
  \author{\authorcr Benno Weck}
  \author{Sara Keen}
  \author{Milad Alizadeh}
  \author{Gagan Narula}
  \author{Matthieu Geist}
  \author{Olivier Pietquin}
  \affil{Earth Species Project}%

\begin{document}

\def\thefootnote{$\star$}\footnotetext{Corresponding author: \url{david@earthspecies.org}}
\def\thefootnote{$\dagger$}\footnotetext{Core authors}
\def\thefootnote{\arabic{footnote}}

\maketitle

\begin{abstract} Large language models (LLMs) prompted with text and audio have achieved state-of-the-art performance across various auditory tasks, including speech, music, and general audio, showing emergent abilities on unseen tasks. 
However, their potential has yet to be fully demonstrated in bioacoustics tasks, such as detecting animal vocalizations in large recordings, classifying rare and endangered species, and labeling context and behavior---tasks that are crucial for conservation, biodiversity monitoring, and animal behavior studies.
In this work, we present \method, the first audio-language foundation model specifically designed for bioacoustics. Our training dataset consists of carefully curated text-audio pairs spanning bioacoustics, speech, and music, designed to address the field's limited availability of annotated data.
We demonstrate successful transfer of learned representations from music and speech to bioacoustics, and our model shows promising generalization to unseen taxa and tasks.
We evaluate \method on a novel benchmark (\newbeans) and it sets a new state of the art on several bioacoustics tasks, including zero-shot classification of unseen species.
To advance bioacoustics research, we release our model weights, benchmark data, and open-source the code for training and benchmark data generation and model training.
\footnote{Project page: \url{https://earthspecies.github.io/naturelm-audio-demo/}}


\end{abstract}

\begin{figure*}[htb]
\begin{center}
\includegraphics[scale=0.23]{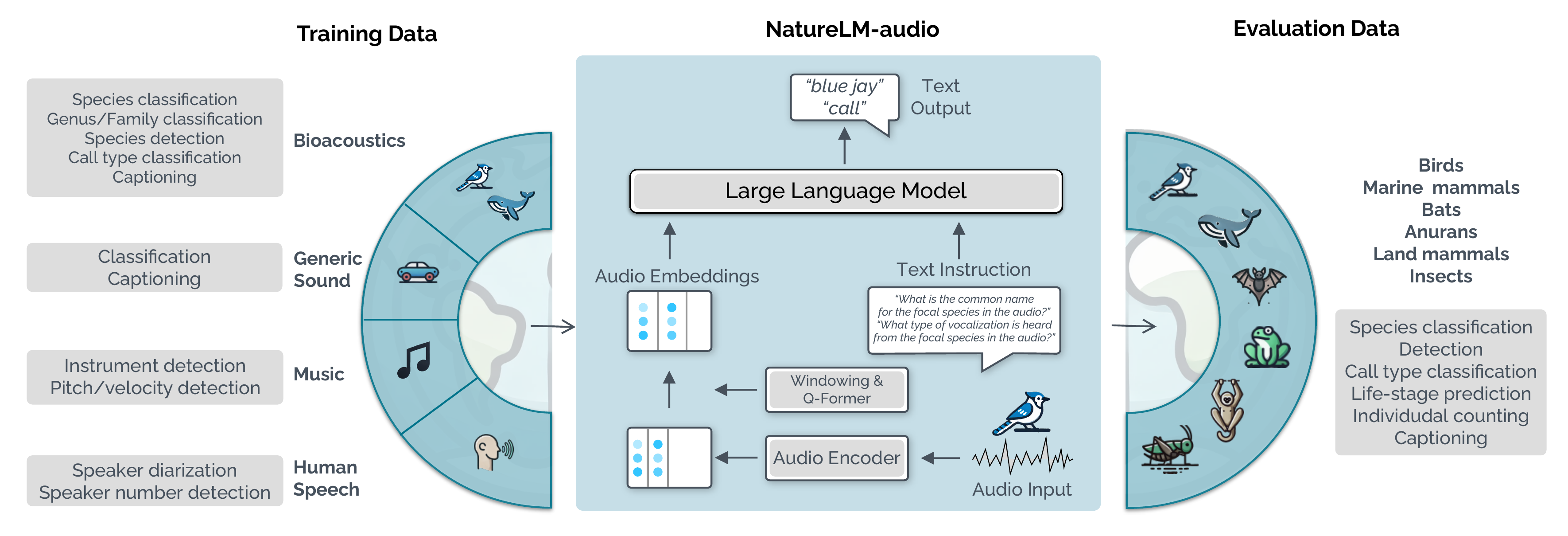} 
\caption{Overview of \method.}
\label{fig:overview}
\end{center}
\end{figure*}

\section{Introduction}
Bioacoustics, the study of sound production and reception in animals, aims to understand animal behavior~\citep{fischer2013bioacoustic}, monitor biodiversity~\citep{stowell2022computational}, and model the mechanisms underlying animal communication~\citep{bradbury1998principles}. 
It plays a vital role in conservation and ecological research, as animal vocalizations provide key insights into ecosystem health, species interactions, and population dynamics. By enabling the detection of endangered species and the tracking of migration patterns, bioacoustic research directly contributes to biodiversity monitoring and conservation efforts~\citep{rutz2023using,stevens2024bioclip}. 

In recent years, machine learning has taken on an increasingly pivotal role in bioacoustic research. Beyond its role in large-scale ecological monitoring, it has opened up new frontiers in the study of animal communication, enabling discoveries such as the use of specialized vocalizations to label conspecifics in marmosets~\citep{marmosets2024}, dolphins~\citep{king2013bottlenose}, and elephants~\citep{pardo2024african}. However, due to inherent challenges in data collection and annotation, many of these studies rely on strongly labeled small datasets~\citep{stowell2022computational} and require careful statistical analysis to ensure significance and mitigate over-fitting. Meanwhile, vast amounts of unannotated bioacoustics data are recorded daily, particularly through passive acoustic monitoring (PAM, ~\cite{dufourq2021automated}) and citizen science platforms such as Xeno-canto~\citep{vellinga2015xeno}. This growing data availability underscores the need for machine learning models capable of large-scale detection, classification, and annotation. The recent successes of large scale AI models in various domains---including natural language processing, computer vision, and game-playing---suggests the possibility of leveraging these large, raw datasets to learn robust and generalizable representations for bioacoustics~\citep{ghani2023global,boudiaf2023search}.

Existing bioacoustics machine learning models are typically designed for specific species or tasks~\citep{dufourq2021automated,kahl2021birdnet,cauzinilleinvestigating}, limiting their generalizability. Many traditional approaches rely on small datasets focusing on a few species and individuals, validating results through statistical measures despite the risks of over-fitting. More recent models, such as BirdNET~\citep{kahl2021birdnet} and Perch~\citep{ghani2023global}, achieve strong performance in bird classification but require training dedicated classifiers for each target taxon. In contrast, we propose a single foundation model that generalizes across taxa.
While recent self-supervised and audio-language contrastive models such as AVES~\citep{hagiwara2023aves} and BioLingual~\citep{robinson2024transferable} have shown promising results on bioacoustics benchmarks, their discriminative and contrastive training paradigms constrain the range of tasks they can effectively address.

In recent years, foundation models---trained on large, diverse datasets, often via self-supervision---have shown promising performance across multiple domains~\citep{bommasani2021on}. While transformer-based large language models (LLMs) are currently the most prominent examples, other architectures, such as diffusion models~\citep{kingma2021variational}, are also emerging as foundation models in some domains. Their ability to handle unseen tasks, perform in-context learning, and respond flexibly to prompts makes them as an appealing alternative to traditional machine learning methods, which typically depend on manually annotated datasets, extensive computational resources, and domain-specific expertise.

While multimodal LLMs, particularly vision-language models (VLMs), have been explored in biodiversity and conservation research~\citep{miao2024new}, large audio-language models (LALMs) remain underexplored for bioacoustics. LALMs have demonstrated strong performance in human speech~\citep{rubenstein2023audiopalm,wang2024retrieval,wu2023decoder,zhang2024speechlm}, music~\citep{gardner2023llark,agostinelli2023musiclm}, and general audio~\citep{tang2024salmonn,chu2024qwen2audiotechnicalreport,gong2023listen}, and they hold significant potential for advancing bioacoustics as well.

%

In this paper, we introduce \method, the first audio-language foundation model specifically designed for bioacoustic tasks, including classification, detection, and captioning. 
Inspired by cross-taxa transfer observed in previous research, such as between human and gibbon or marmosets~\citep{cauzinilleinvestigating, sarkar2023can} and birds and whales~\citep{ghani2023global}, we incorporate speech and music tasks into training. We show that representations learned from these domains successfully transfer to animal vocalizations, demonstrating generalization across species.
Additionally, we expand the BEANS bioacoustics benchmark~\citep{hagiwara2023beans} with new tasks, including call-type prediction, lifestage classification, captioning, and individual counting. This new benchmark, \newbeans, enables us to evaluate cross-domain learning and zero-shot transfer to unseen taxa and tasks. Unlike existing bioacoustics benchmarks such as Perch (\cite{ghani2023global} for bird detection) and BirdSet (\cite{rauch2025birdset} for bird classification), we do not focus solely on birds and we go beyond species classification. Additionally, our dataset presents prompts and audio descriptions in natural language, fostering further research in LALMs.
\looseness=-1


Our contributions are as follows:
{\bf (i) Model}: We introduce \method, the first audio-language foundation model for bioacoustics, trained on a carefully curated dataset spanning animal vocalizations, general audio, human speech, and music.
{\bf (ii) Domain transfer}: We show that our model generalizes beyond the species seen during training and exhibits zero-shot capabilities on unseen taxa and species. 
{\bf (iii) Task transfer}: We evaluate our model on \newbeans, which extends beyond species classification and includes unseen tasks such as individual counting. For the first time, we show positive transfer from speech and music data to bioacoustics tasks.


\section{Related Work}

Most prior work on audio-language models has focused on human speech processing. Models such as SpeechGPT \citep{zhang2023speechgpt}, Speech-LLaMA \citep{wu2023decoder}, AudioLM \citep{borsos2023audiolm}, AudioPaLM \citep{rubenstein2023audiopalm}, AudioGPT \citep{huang2023audiogpt}, SpiRit-LM \citep{nguyen2024spiritlm}, and SpeechLM \citep{zhang2024speechlm} mostly focus on building language models that can perceive and produce human speech.
While such models could, in principle, be fine-tuned for bioacoustic tasks, doing so would require significant computational resources and domain expertise. Instead, our model shows promising generalization to unseen species and tasks without requiring additional fine-tuning.
\looseness=-1

Recently, more general language models with audio perception capabilities have emerged. Pengi~\citep{deshmukh2023pengi} uses an audio encoder and a text encoder mapped onto an LLM to perform audio-to-text tasks. SALMONN~\citep{tang2024salmonn} uses dual audio encoders and integrates Q-Former~\citep{li2023blip2} to improve the handling of speech and general audio inputs. Qwen-audio~\citep{chu2023qwenaudio} adopts a multi-task learning approach with the introduction of the Speech Recognition with Timestamp (SRWT) task. LTU~\citep{gong2023listen} builds an open-ended question-answer dataset and applies curriculum learning strategies to improve generalization. Similar multimodal models have been proposed for music, such as MU-LLaMA~\citep{liu2023music} and LLark~\citep{gardner2023llark}.
While recent foundation models such as AVES~\citep{hagiwara2023aves} and BioLingual~\citep{robinson2024transferable} have demonstrated promising results in bioacoustics, their training paradigms and architectures constrain the range of tasks they can address.

Although animal sounds and vocalizations are often part of generic audio datasets such as AudioSet~\citep{gemmeke2017audioset} and audio caption datasets~\citep{kim2019audiocaps,mei2023wavcaps}, these datasets are often too broad and lack the fine-grained annotations required for bioacoustic tasks such as species classification, behavior analysis, or ecological monitoring. As a result, LALMs trained on these datasets tend to produce only generic labels (e.g., `bird' rather than a specific species name). We address this limitation by introducing a diverse, multi-task training dataset and \method, an LALM designed to produce robust representations for bioacoustics.

While specific bioacoustics benchmarks such as BIRB~\citep{hamer2023birb} for bird vocalization retrieval and BEANS~\citep{hagiwara2023beans} for classification and detection exist, the field still lacks comprehensive benchmarks comparable to those in human speech and music, such as Dynamic-SUPERB~\citep{huang2024dynamicsuperb} or AIR-Bench~\citep{yang2024airbench}. This gap limits the evaluation of bioacoustic models, particularly in areas such as zero-shot learning and task generalization.

In this work, we aim to bridge these gaps by introducing \method, the first audio-language foundation model specifically designed for bioacoustics, and \newbeans, an expanded benchmark that evaluates cross-species and cross-task generalization.

\section{Training Dataset Creation}
\label{sec:training_data}

\begin{table}[!t]
\begin{center}
{\small
\begin{tabular}{@{}llrrl@{}}
\toprule
Task$^a$        &     Dataset     & \# Hours & \# Samples \cmmnt{& \# Labels}\\ \midrule
CAP     & {\tt WavCaps} \citep{mei2023wavcaps}        & 7568        & 402k  \cmmnt{& description, species}  \\
CAP           & {\tt AudioCaps} \citep{kim2019audiocaps}    & 145      & 52k   \cmmnt{&   description, species }  \\
CLS          & {\tt NSynth} ~\citep{engel2017neural}        & 442        & 300k   \cmmnt{&  instruments, pitch}\\
CLS          & {\tt LibriTTS} \citep{zen2019libritts}, {\tt VCTK} \citep{yamagishi2019cstr}      & 689        & 337k  \cmmnt{& n. speakers}\\
CAP        & {\tt Clotho}~\citep{drossos2020clotho}         &  25     &  4k    \\
CLS, DET, CAP    &   {\tt Xeno-canto} ~\citep{vellinga2015xeno}         & 10416     & 607k   \cmmnt{& species, call, age} \\
CLS, DET, CAP     &  {\tt iNaturalist} ~\citep{inaturalistSite}         & 1539         & 320k    \cmmnt{& species} \\
CLS, DET, CAP     &   {\tt Watkins} ~\citep{sayigh2016watkins}        & 27         & 15k        \cmmnt{& species} \\ 
CLS, DET     & {\tt ASA} ~\citep{AnimalSoundArchiveAccess}        & 78           & 16k        \cmmnt{& species} \\
DET     &  {\tt Sapsucker Woods} ~\citep{kahl2022ssw}        & 285         & 342k        \cmmnt{& species} \\ 
CLS, DET     &  {\tt Barkley Canyon} ~\citep{Barkley}        & 876         & 309k        \cmmnt{& species} \\ 

  CLS   & {\tt Urbansound} ~\citep{salamon2014dataset}          & 10        & 2k       &  \\
     \bottomrule
\end{tabular}
}
\end{center}
\caption{Training tasks and datasets. $^a$ CLS: classification, DET: detection, CAP: captioning.}
\label{table:training_datasets}
\end{table}

\begin{figure*}[htb]
\begin{center}
\includegraphics[scale=0.40]{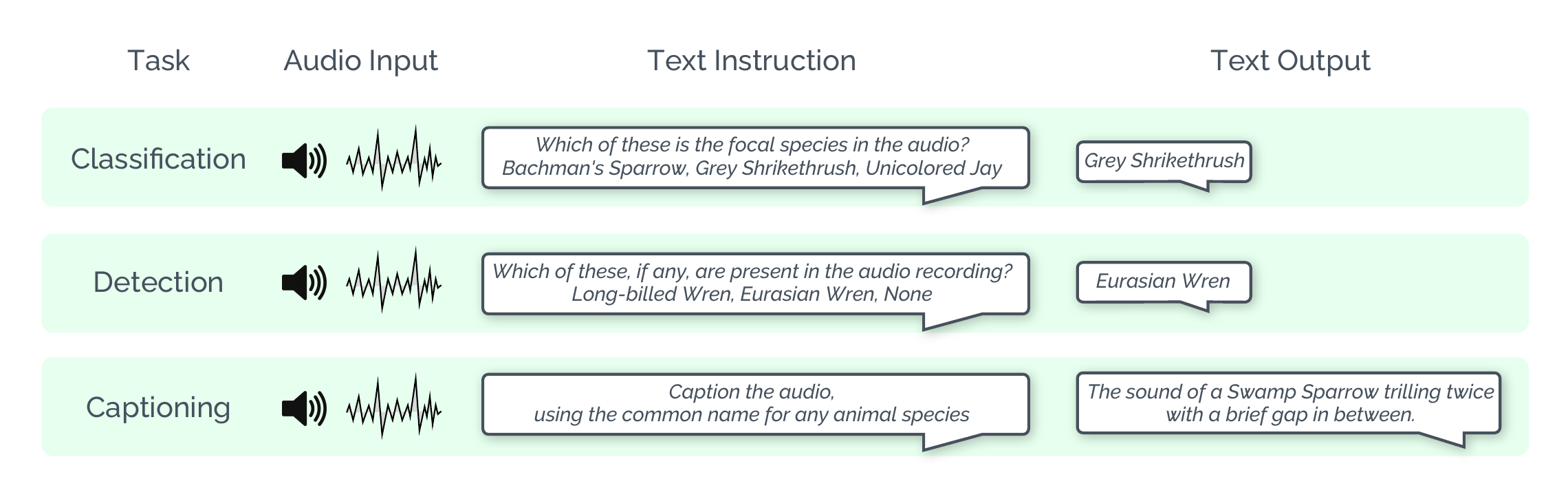} 
\caption{Examples of training instances.}
\label{fig:examples}
\end{center}
\end{figure*}
To train an audio-text model for bioacoustics, we compile a diverse dataset of text-audio pairs (Table~\ref{table:training_datasets}). The data is collected through a combination of prompting on existing audio datasets, generating new LLM-generated text labels, and mixing new, procedurally augmented audio data. The dataset is comprised of bioacoustic recordings, general audio, speech, and music datasets. Figure~\ref{fig:examples} shows examples of training instances used for \method. We plot the distribution of the training samples in Figure \ref{fig:training_data} in the Appendix.

\subsection{Bioacoustic Data}

\textbf{Species Classification}: 
We standardize large-scale bioacoustic archives into a common format, processing datasets such as Xeno-canto \citep{xenocanto}, iNaturalist \citep{inaturalistSite}, Animal Sound Archive \citep{AnimalSoundArchiveAccess}, and Watkins (all-cuts, \cite{sayigh2016watkins}). Differences in species naming conventions across datasets are reconciled using the GBIF taxonomy backbone \citep{gbif_taxonomy_backbone}. We prompt the model to predict the scientific name, common name, or ``taxonomic name'' of the focal species or all species present in a recording. Taxonomic names are written as ``phylum class order family genus species'' and are inspired by BioCLIP \citep{stevens2024bioclip}, which found that flattening the hierarchy into a text name improved generalization to unseen species in computer vision. In many real-world applications, an animal vocalization is known to belong to a subset of species---for example, based on geographic location. To model this, we generate prompts that present the model with a set of candidate species as possible answers. For 30\% of prompts, we sample ``random'' negatives by selecting from all common names or scientific names in our dataset. In the remaining prompts, we introduce ``hard'' negatives by selecting species that share a common ancestor at the family, order, or phylum level. The number of negative samples is randomly selected, up to a maximum of 35.

Unlike traditional bioacoustic models that predict based on audio alone, the text-audio formulation enables classification conditioned on additional context. We train the model to classify species while conditioned on recording metadata and field notes. We follow the same setup as above, but inject the time of the recording, the location, and the free-text notes of the recordist into the prompt. This data is added wherever available for Xeno-canto, with time, location, or field note components randomly dropped a percentage of the time.

To avoid data leakage, we exclude a set of held-out species and the cbi and Watkins data used in \newbeans.

\textbf{Species Detection}: 
Using the same datasets as in species classification, we prompt the model to determine whether a given species is present in a recording. The model selects from a provided set of candidate species or chooses ``None'' when no correct option is given. Candidate sets are constructed with a mix of random and hard negatives, similar to the classification task. In 50\% of prompts, the correct species is omitted from the set, making ``None'' the correct answer.
\looseness=-1



Because Xeno-canto comprises mostly focal recordings, we account for the covariate shifts in soundscapes by adding noise---audio that does not contain animal vocalizations, speech, or music. We source noise samples from datasets including: ShipsEar~\citep{santos2016shipsear}, Deepship~\citep{irfan2021deepship} and Orcalab~\citep{poupard2020massive} for boat engine sounds, as well as FSD50K~\citep{fonseca2021fsd50k} and Urbansound~\citep{salamon2014dataset} for non-animal, non-music sound classes, and all the classes from TUT2016~\citep{mesaros2016tut}, IDMT~\citep{abesser2021idmt}, Demand~\citep{thiemann2013diverse}, and Wham noise~\citep{wichern2019wham}. 
The noise is added programmatically, using random files at a random signal-to-noise ratio (SNR) sampled from a uniform distribution between $-10$dB and $20$dB.

In addition, we used soundscape recording datasets for detection from Sapsucker Woods (SSW \cite{kahl2022ssw}) for birds and from Barkley Canyon \citep{Barkley, ONC2013, ONC2014a, ONC2014b} for marine mammals. Following the BEANS detection dataset methodology, we segment audio into 10-second windows with a 5-second overlap, and treated it as a multi-label classification problem. Species with more than 100 occurrences were used as target labels, while those with fewer occurrences were grouped into an ``other'' class.

\textbf{Captioning}: 
For bioacoustic captioning, we use the AnimalSpeak dataset~\citep{robinson2024transferable}, which aggregates bioacoustic datasets into a language-model-captioned dataset. We add separate prompts for captioning with scientific vs. common names, for ``rich'' captions over eight words, and for templated captions from Xeno-canto which follow a strict structure.

\textbf{Call-type and Lifestage}: 
We include multiple new bioacoustic tasks which can be expressed based on the Xeno-canto metadata. Specifically, predicting the life stage of birds, predicting call-types, and differentiating between calls and songs. The model is prompted using either audio alone or audio with the species name. Additionally, we include marine mammal call-type classification using Barkley Canyon recordings. These tasks go beyond species classification, providing finer-grained insights into ecological monitoring and animal behavior studies.



\subsection{Non-bioacoustic Data}
%
\paragraph{General Audio} We include WavCaps~\citep{mei2023wavcaps}, AudioCaps~\citep{kim2019audiocaps}, and Clotho~\citep{drossos2020clotho} for general audio captioning. We observe that, during WavCaps creation, some recordings originally contained metadata relevant to bioacoustics and specific species. However, this metadata was lost during general-domain captioning, resulting in overly generic descriptions. We identify such cases by analyzing the original metadata, and re-process the metadata prompting Gemini-1.0-pro to produce bioacoustic captions. These enhanced captions are included alongside the original ones.



\paragraph{Music} Pitch, timbre, and the number of animals in a recording are key acoustic features used by biologists to infer context and behavior.
We use NSynth 2.3.3~\citep{engel2017neural} to create a set of tasks that may help bioacoustics downstream tasks.
We generate text prompts for \textit{pitch detection} in Hz, \textit{instrument name}, and \textit{velocity}, ranging 0 to 1. Additionally, we use the timbre `qualities' labels to create \textit{text descriptions} for each audio. For instance, if the sound is `distorted,' we generate descriptions such as ``This sound has a distinctive crunchy sound and presence of many harmonics.'' or ``This sound is distorted''. 
Moreover, we create synthetic mixtures by layering one to three different instruments. In this case we generate two tasks: predicting the \textit{number of instruments} and identifying the \textit{instrument names}.

\paragraph{Speech} We use LibriTTS~\citep{zen2019libritts} and VCTK~\citep{yamagishi2019cstr} to generate synthetic mixtures of up to four speakers, a task that may transfer to individual counting in bioacoustics. To better match the frequency variability in animal vocalizations, we time-scale the speech mixtures with factors sampled from an uniform distribution between 0.25 to 4 (i.e., from 4x slower to 4x faster). Since animal vocalizations tend to be sparse, we insert random segments of silence at local minima computed on the RMS of the speech signals. To enhance realism, we further convolve the generated mixtures with impulse responses sampled from the DNS Challenge~\citep{dubey2024icassp}.

\section{Evaluation data: the \newbeans benchmark}

\begin{table}[!t]
\begin{center}
{
\begin{tabular}{@{}lllrl@{}}
\toprule
Task$^a$        &     Dataset         & Description             & \# Size$^b$ & \# Labels (type) \\ \midrule
CLS     & {\tt esc50}          & generic sound            & 400        & 50 (sound type)  \\
CLS          & {\tt watkins}         & marine mammals            & 339        & 31 (species)     \\
CLS        & {\tt cbi}             & birds                   &  3620     &  264 (species)     \\
CLS     &   {\tt humbugdb}        & mosquito                & 1859      & 14 (species)     \\
DET     &  {\tt dcase}           & birds \& mammals          & 13688         & 20 (species)  \\
DET     & {\tt enabirds}        & birds                   & 4543           & 34 (species)     \\
DET     &  {\tt hiceas}          & cetaceans               & 1485         & 1 (species)      \\
DET     & {\tt rfcx}            & birds \& frogs          & 10406        & 24 (species)     \\
DET     & {\tt gibbons}         & gibbons                 & 18560         & 3 (call type)    \\ \midrule
CLS     & {\tt unseen-species}       & birds etc.             & 1255         & 200 (species)     \\
CLS     & {\tt unseen-genus}       & birds etc.             & 951         & 101 (species)     \\
CLS     & {\tt unseen-family}       & birds etc.             & 451         & 36 (species)     \\
CLS     & {\tt lifestage}         & birds                   & 466         & 3 (stage)        \\
CLS     & {\tt call-type}         & birds                    & 1000         & 2 (call/song)  \\
CAP     & {\tt captioning}        & birds etc.                & 29002         & (open-ended)      \\ 
CLS     & {\tt zf-indv}           & zebra finches            & 1160         & 2 (\# of indv.)   \\ \bottomrule
\end{tabular}
}
\end{center}
\caption{Evaluation tasks and datasets of \newbeans. $^a$ CLS: classification, DET: detection, CAP: captioning. $^b$ The numbers of samples for classification and captioning, and the number of 5-second ``chunks'' for detection (see Section~\ref{sec:training_data} for more details). \looseness=-1}
\label{table:datasets}
\end{table}

One of the key contributions of this work is \newbeans, a new benchmark for bioacoustics (Table~\ref{table:datasets}). \newbeans extends beyond traditional species classification by introducing new tasks such as call-type prediction, lifestage classification, captioning, and individual counting, which is not seen during training. To construct \newbeans, begin with the test portion of BEANS~\citep{hagiwara2023beans}  evaluates models on standard bioacoustic tasks and datasets, including:
\looseness=-1
\begin{itemize}
\setlength\itemsep{0.1em}
    \item {\tt esc50}~\citep{piczak2015dataset}: Generic environmental sound classification with 50 labels.
    \item {\tt watkins}~\citep{sayigh2016watkins}: Marine mammal species classification with 31 species.
    \item {\tt cbi}~\citep{cornell2020}: Bird species classification with 264 labels from the Cornell Bird Identification competition hosted on Kaggle.
    \item {\tt humdubdb}~\citep{kiskin2021humbugdb}: Mosquito wingbeat sound classification into 14 species.
    \item {\tt dcase}~\citep{morfi2021fewshot}: Mammal and bird detection from DCASE 2021 Task 5: Few-shot Bioacoustic Event Detection (20 species).
    \item {\tt enabirds}~\citep{chronister2021annotated}: Bird dawn chorus detection (34 species).
    \item {\tt hiceas}~\citep{noaa2022hawaiian}: Minke whale detection from the Hawaiian Islands Cetacean and Ecosystem Assessment Survey (HICEAS) (1 label).
    \item {\tt rfcx}~\citep{lebien2020pipeline}: Bird and frog detection from the Rainforest Connection (RFCx) data with 24 species.
    \item {\tt gibbons}~\citep{dufourq2021automated}: Hainan gibbon detection with 3 call type labels.
\end{itemize}
We also include novel bioacoustics datasets including:
\begin{itemize}
\setlength\itemsep{0.1em}
    \item {\tt unseen-species}: 200 species held out from AnimalSpeak \citep{robinson2024transferable}. For a controlled measure of generalization, we hold out species whose genus is well-represented (at least 150 training examples)
    \item {\tt unseen-genus}: We hold out entire genus whose family is well-represented (at least 250 training examples) totaling 101 unique species.
    \item {\tt unseen-family}: We hold out entire families whose class is well-represented (at least training 250 examples) totaling 36 unique species and representing the hardest generalization setting.
    \item {\tt lifestage}: Predicting the lifestage of birds across multiple species. Newly curated from Xeno-canto~\citep{xenocanto}.
    \item {\tt call-type}: Classifying song vs. call across multiple bird species. Newly curated from Xeno-canto~\citep{xenocanto}.
    \item {\tt captioning}: Captioning bioacoustic audio on AnimalSpeak~\citep{robinson2024transferable}.
    \item {\tt zf-indv}~\citep{elie2016vocal}: Determining whether a recording contains multiple zebra finches, using programmatically generated mixtures (1–4 individuals).
\end{itemize}
Some of these tasks, particularly bioacoustic captioning, have not been extensively studied before. Captioning allows for automatic generation of descriptive annotations of animal sounds, enhancing our understanding of species behaviors and communication patterns. Improvements in other new tasks, such as cross-species lifestage and call-type prediction, would allow finer-grained ecological monitoring and animal communication studies at scale.

For evaluation, we use accuracy for classification, macro-averaged F1 for detection, and SPIDEr~\citep{liu2017spider} for captioning. Unlike mean average precision (mAP), which is originally used in BEANS and assumes a smooth ranking of candidates, F1 is more appropriate for evaluating generative tasks. This ensures a fairer assessment of models that generate predictions instead of ranking pre-defined classes.

\section{\method  Architecture}

Our model follows a generic audio-to-text architecture similar to prior LALMs such as SALMONN~\citep{tang2024salmonn}, Qwen2-audio~\citep{chu2024qwen2audiotechnicalreport}, and LTU~\citep{gong2023listen}. These models are trained on paired audio-text data for tasks including speech, music, and general audio event understanding. Figure~\ref{fig:overview} provides an overview of the \method architecture.

\method first encodes the input audio using BEATs~\citep{chen2023beats}, a state-of-the-art audio encoder on multiple audio tasks. To connect the BEATs embeddings with the LLM, we use a Q-Former~\citep{li2023blip2} applied at the window level as proposed in SALMONN~\citep{tang2024salmonn}. 
Similarly to the existing LALMs, we use an LLM to produce text, in this case Llama 3.1-8b~\citep{dubey2024llama3}, which is fine-tuned with LoRA~\citep{hu2022lora}. During training, only the adapter layers of the LLM are updated, while the base LLM parameters remain frozen. In contrast, the audio encoder and Q-Former remain trainable.
The model takes an audio input $\va$ along with an instruction $\vx$ and produces a text output $\vy$.
The model is trained under the loss function:
\begin{eqnarray}
    \vh &=& f_W({\rm Encoder}(\va)) \\
    \vz &=& p_\varphi^Q(\vq, \vh) \\
    L &=& -\sum \log p_\theta^{LM}(\vy_{t} |\vx, \vz, \vy_{<t})
\end{eqnarray}
where ${\rm Encoder}$ is the pretrained BEATs~\citep{chen2023beats} audio encoder, $f_W$ is a function that converts consecutive $W$ audio frames into a window, $p_\varphi^Q$ is the Q-Former model with trainable parameters $\varphi$ that converts a window into a sequence of text representations $\vz$ using query $\vq$, and $p_\theta^{LM}$ is the pretrained LLM with trainable parameters $\theta$.

\section{Training Method}
Our training method follows a curriculum learning approach~\citep{soviany2021curriculum}, where the model is first trained on simpler tasks before progressively tackling more complex ones, as done in other audio foundation models~\citep{tang2024salmonn, gong2023listen}. We train in the two stages:
\begin{itemize}
\item Stage 1 (Perception Pretraining):
We pretrain the model exclusively on focal species classification, classifying vocalizations from thousands of animal species. Species classification is a highly deterministic task, allowing opportunity to learn a robust connection between language and audio. We also choose to train on this task individually as it is foundational to other tasks in bioacoustics.
\item Stage 2 (Generalization Fine-tuning):
In the second stage, we introduce a variety of bioacoustic and other tasks, building on the robust classification abilities developedin Stage 1. This includes detection, captioning, lifestage prediction, and call-type prediction. We also include speech and music data in this second stage, aimed at improving transfer to bioacoustic tasks.
%
\end{itemize}
We train \method from scratch, initializing the Q-Former and LoRA layers randomly rather than fine-tuning existing LALM checkpoints such as SALMONN. This allows for more flexibility in terms of choosing the latest LLM with the extensive knowledge of animal species, and the most relevant architectural components (e.g., excluding memory-intensive parts of current LALMs such as the Whisper speech encoder~\citep{radford2022whisper}). 

\section{Experiments}
\subsection{Training and Evaluation Details}

We train our model on the full curated training set (Section \ref{sec:training_data}). 
To evaluate generalization, we create hold-out splits for Xeno-canto, iNaturalist, Animal Sound Archive, and Watkins datasets, used solely for benchmarking.

We initialize the audio encoder weights using an existing BEATs checkpoint\footnote{\tt BEATs\_iter3\_plus\_AS2M\_finetuned\_on\_AS2M\_cpt2.pt} and fully fine-tune it, which we found to be critical in an ablation (Table \ref{tab:freeze-beats}). We initialize the LLM from Llama-3.1-8B-Instruct and apply LoRA to all attention layers (rank: 32, alpha: 32, dropout: 0.1).

We follow the proposed two-stage training strategy. In both stages, we use a linear warmup followed by a cosine learning rate schedule, with a peak learning rate of $9.0 \times 10^{-5}$ and an end learning rate of $2.0 \times 10^{-5}$. We use a batch size of $128$ and run the first stage for $5.0 \times 10^{5}$ steps and the second stage for $1.6 \times 10^{6}$ steps. For inference, we use beam search with two beams, a repetition penalty of 1.0, and a length penalty of 1.0.

We consider several inference methods depending on the task type. Species-classification tasks involve single-label prediction: we prompt the model to output the species name from the recording. Since the LLM may generate text that does not exactly match predefined labels, we use Levenshtein distance to map predictions to the closest species name. We choose the Levenshtein distance for its simplicity and because species names, in particular Latin names, have high character-overlap with related names. However, we note that it may not be optimal for general audio classification.
\looseness=-1

For multilabel detection tasks, the number of target species varies by dataset. For tasks with 10 or fewer species, we include the species options in the prompt. Otherwise we prompt the model to list all species in the audio, if any. In both cases, the model outputs all detected species, or `None'. We discard predictions with low character overlap with the valid labels.

Our baselines include CLAP-like models~\citep{wu2023clap}, which cannot naively perform multilabel detection. To address this, we create a negative ``template'' for each detection task, as proposed by~\citet{miao2023zeroshot}. We consider each label a detection positive for CLAP if the audio is more similar to the label than to the negative template in the CLAP model's embedding space.

\subsection{Species Classification and Detection}

\begin{table}[!t]
\begin{center}
{
\begin{tabular}{lrrrrrrrrr}
\toprule
         Model & esc50 &  watkins &   cbi &  humbugdb &  dcase &  enabirds &  hiceas &   rfcx &   gibbons \\
\midrule
 LLM w/o audio & 0.020 & 0.041 & 0.005 & 0.073 &  0.000 & 0.001 &   0.210 & 0.000 & {\ul 0.013} \\
       SALMONN &   0.320 & 0.041 & 0.004 & {\ul 0.090} &  0.005 & 0.004 &   0.097 & 0.002 & 0.005 \\
    Qwen2-audio &   0.307 & 0.041 & 0.004 & 0.070 &  0.005 & 0.004 &   0.097 & 0.002 & 0.005 \\
    BioLingual & {\ul 0.600} & {\ul 0.257} & {\ul 0.705} & 0.085 &  {\ul 0.036} & {\ul 0.109} &   {\bf 0.429} & {\ul 0.004} & {\bf 0.018} \\
\method & {\bf 0.820} & {\bf 0.788} & {\bf 0.778} & {\bf 0.114} &  {\bf 0.058} & {\bf 0.314} &  {\ul 0.336} & {\bf 0.025} & 0.005 \\
\bottomrule
\end{tabular}
}
\end{center}
\caption{Main zero-shot results on \newbeans. We used accuracy for classification, and F1 for detection tasks. The best and the second best metrics are highlighted and underlined per each dataset.}
\label{tab:beans}
\end{table}

Table~\ref{tab:beans} shows the main results measured on the \newbeans species classification and detection datasets. Our baselines include an LLM (the original Llama-3.1-8B-Instruct model without fine-tuning, \cite{dubey2024llama3}) without audio input, SALMONN~\citep{tang2024salmonn}, BioLingual~\citep{robinson2024transferable}, and Qwen2-audio~\citep{chu2024qwen2audiotechnicalreport}. All baselines are evaluated in the same way as \method. As shown in the table, the outputs from the LLM without audio input, SALMONN, and Qwen2-audio are largely random on bioacoustic datasets, failing to properly interpret the input audio or follow the instructions. In contrast, \method achieved state-of-the-art zero-shot performance on 7 out of 9 datasets, and delivered competitive results on the remaining tasks from the \newbeans benchmark. 
We note that performance of baselines on the general audio dataset ESC50~\citep{piczak2015dataset} may be reduced by the use of the Levenshtein distance, as our pipeline is optimized for bioacoustic tasks.
\looseness=-1

\begin{table}[!t]
\begin{center}
{
\begin{tabular}{lrrr}
\toprule
       & cbi   & dcase-bird & enabirds  \\ \midrule
BirdNET     &    0.609       &    0.035       &   0.490   \\
Perch         &    0.744       &     0.035      & 0.164     \\
\method      &   0.778     &     0.083      &   0.314  \\
\bottomrule

\end{tabular}
}
\end{center}
\caption{Comparison with bird vocalization models.}

\label{tab:birds}
\end{table}

We also compared \method~with bird-specific classification models, namely BirdNET~\citep{kahl2021birdnet} and Perch~\citep{ghani2023global}, to evaluate the zero-shot capabilities of our model. We compare on the bird-related datasets of \newbeans, plus the portion of DCASE with bird species. Results are presented in Table~\ref{tab:birds}. 
Since both BirdNET and Perch were trained in a supervised manner on datasets that significantly overlap with our bird evaluation datasets, this is not a fully fair comparison, and their performance should be considered as topline results. Nevertheless, our model demonstrated strong zero-shot bird vocalization classification capabilities. In particular, we achieve a new SotA for the cbi dataset, classifying vocalizations of hundreds of birds, and achieve competitive results with the bird-specific models on both detection tasks. We additionally compare against various models on datasets from the BirdSet benchmark (\cite{rauch2025birdset}, where our model achieves the highest average top-1 accuracy (Appendix in Table \ref{tab:t1-acc-only}).

\subsection{Generalizing to Unseen Species}

We further evaluate the model’s ability to generalize to completely unseen taxa using the newly added datasets in \newbeans, held out at three levels: unseen species, unseen genus, and unseen families. As a topline, we compare against BioLingual, which has seen these taxa in training and only indicates fully supervised performance. As baselines, we consider a theoretical random baseline (1 / number of classes) and CLAP-LAION~\citep{elizalde2023clap}, a general-domain audio model which, similar to our model, is unlikely to have seen these species during training. We compare the performance when predicting common, scientific, or taxonomic names. 

Table~\ref{tab:unseen} presents the results. Across all three unseen taxa settings, \method significantly outperforms the random baseline, demonstrating its ability to generalize to unseen taxa and taxonomic branches. For example, on the unseen species test set, our model achieves an accuracy of 34.3\%, far surpassing the random baseline of 0.5\%, indicating that the model has learned features that extend beyond the species it was trained on. The model also outperforms CLAP-LAION, further emphasizing its ability to generalize. We observe that predicting with taxonomic names consistently improves performance across all settings, and is particularly critical for generalizing to unseen genus and families where scientific (Latin) names alone fail to capture hierarchical relationships. We further note that scientific names perform relatively well when generalizing to unseen species, but perform worse than common names for generalizing to unseen genus, This suggests that common names may encode broader hierarchical information or be more familiar to the language model.

\begin{table*}[!t]
\begin{center}
{
\begin{tabular}{lrrrr}
\toprule
       & unseen-species$^a$   & unseen-genus$^b$  & unseen-family$^c$   \\ \midrule    
Supervised SotA    & 0.687     &   0.688      & 0.545   \\ \midrule
random chance    &      0.005   &  0.010      & 0.028   \\ 
baseline (CLAP)    &      0.014   &  0.026      & 0.082   \\ 
\method (cmn)       &      0.181   &  0.116      & 0.035    \\
\method (sci)       &      0.238   &  0.041      & 0.035   \\ 
\method (tax)      &      \textbf{0.343}   &  \textbf{0.148}      & 
 \textbf{0.308}   \\ \bottomrule

\end{tabular}
}
\end{center}
\caption{Generalization to unseen taxa in terms of classification accuracy. All tasks predict species names, on test sets held-out at the $^a$ species $^b$ genus and $^c$ family level. Targets were not held out from ``Supervised SotA'' reference (BioLingual). Cmn, sci, and tax denote predictions using common, scientific, and taxonomic names respectively. Since the number of labels varies across datasets, results should not be directly compared across columns.}
\label{tab:unseen}
\end{table*}

\subsection{Novel Bioacoustic Tasks}

Beyond species classification, we evaluate \method on novel bioacoustic tasks introduced in \newbeans, which, to the best of our knowledge, have not been previously studied at a cross-species level. We additionally include {\tt zf-indv}, a completely unseen task that determines whether a recording contains multiple zebra finch individuals or just one~\citep{elie2016vocal}. We compare against BioLingual~\citep{robinson2024transferable} for discriminative tasks and SALMONN~\citep{tang2024salmonn} for captioning. As shown in Table~\ref{tab:noveltasks}, \method sets a new state-of-the-art across all tasks.
We evaluate call-type classification more extensively (Table \ref{tab:call-type-transfer}), and find the model is able to transfer this task to unseen taxa. We further find the model can improve audio classification performance by incorporating additional context as text, which we discuss in the Appendix in \ref{app:species_context}.

\begin{table}[!t]
\begin{center}
{
\begin{tabular}{lrrrr}
\toprule
       & lifestage   & call-type &  captioning  & zf-indv \\ \midrule
 SotA          &    0.502       &   0.658       &   0.009    & 0.604  \\
\method       &     0.794      &      0.871    &    0.532   &  0.655  \\ 
\bottomrule

\end{tabular}
}
\end{center}
\caption{Results on \newbeans novel bioacoustics tasks. We report accuracy for classification, and SPIDEr~\citep{sharif2018learning} for captioning. SotA is SALMONN for captioning and BioLingual for the remaining tasks.}
\label{tab:noveltasks}
\end{table}

\subsection{Ablation on Speech and Music}
To investigate the impact of speech and music on downstream task performance, we run an ablation during stage-2 training. Specifically, we train two versions of the model for 150k steps---one with speech and music data and one without---and evaluate their ability to perform an unseen task: counting zebra finches. The model trained with speech achieves $67.7\%$, similar to our full model. The model trained without speech scored $50.0\%$, exactly random, and qualitatively predicted `more than one' for all examples. These results suggest the ability to count vocalizing birds transfers from human speech and music, as our training data includes tasks such as counting human speakers in a recording. We include the ablation performance on all tasks in the Appendix (Tables~\ref{tab:speech-beans} and~\ref{tab:speech-added}).

\section{Conclusion}
We presented \method, the first audio-language foundation model specifically designed for bioacoustics, demonstrating its potential to address critical tasks such as classifying and detecting animal vocalizations, and decoding context, call types, and individuals across species. By leveraging a carefully curated dataset spanning bioacoustics, speech, and music data, \method sets the new state-of-the-art on multiple tasks, including zero-shot classification of unseen species. Moreover, our model demonstrates positive transfer across both domains and tasks, performing well on a novel benchmark (\newbeans), which includes new bioacoustic tasks such as captioning and individual counting. To further accelerate research and the development of more robust models in the field, we have open-sourced the code for generating both training and benchmarking data.

We plan to extend this work by incorporating more diverse tasks and datasets, improving the text-based LLM backbone with bioacoustic-specific texts, and enhancing the model's multilingual capabilities. Another direction is the introduction of new modalities, such as motion and image data, leading to multimodal models like NatureLM-motion and NatureLM-image. We also aim to explore the model's generative abilities, particularly in producing audio tokens for applications such as animal sound synthesis and audio denoising.

While \method offers significant potential for advancing biodiversity monitoring and conservation, several ethical concerns must be addressed. First, there is a potential bias towards bird vocalizations due to the overrepresentation of bird datasets, which could limit the model’s effectiveness in other taxa. Second, the model’s ability to detect and classify endangered species could be misused for illegal activities such as poaching, posing a threat to wildlife. Finally, unintended consequences on animal behavior and ecology must be considered, particularly when deploying LLMs, known for their issues including hallucinations and biases~\citep{kuan2024understanding}. These systems may interfere with the behavior of the species being studied, and the long-term ecological impact of widespread passive monitoring is still unknown. Careful deployment and responsible use are essential to mitigate these risks.
%
%
\bibliography{references}
\bibliographystyle{iclr2025_conference}

\newpage
\appendix


\startcontents[appendixmaterialtoc]

\printcontents[appendixmaterialtoc] 
                 {l}               
                 {1}{              
                   \setcounter{tocdepth}{2} 
                 }

\section{Held-out data}
\subsection{Held-out Families}
\begin{multicols}{3}
\begin{enumerate}
\setlength\itemsep{0.0em}
\tiny
\item Elachuridae
\item Calyptophilidae
\item Pelecanoididae
\item Phocoenidae
\item Alytidae
\item Castoridae
\item Dicroglossidae
\item Suidae
\item Prophalangopsidae
\item Octodontidae

\end{enumerate}
\end{multicols}

\subsection{Held-out Genus}
\begin{multicols}{3}
\begin{enumerate}
\setlength\itemsep{0.0em}
\tiny

\item Aglaeactis
\item Drepanorhynchus
\item Lesbia
\item Nemobius
\item Meconema
\item Pseudochorthippus
\item Caliechthrus
\item Pachycare
\item Rhodothraupis
\item Astrapia
\item Probosciger
\item Amazonetta
\item Ocyalus
\item Nandayus
\item Rhinocrypta
\item Heterocercus
\item Jacamaralcyon
\item Hymenops
\item Doliornis
\item Eugerygone
\item Cryptosylvicola
\item Taeniopygia
\item Catharopeza
\item Eurostopodus
\item Tylas
\item Vini
\item Ptychoramphus
\item Speculanas
\item Aphelocephala
\item Stipiturus
\item Procarduelis
\item Rhopophilus
\item Neopsephotus
\item Enodes
\item Leucocarbo
\item Gymnophaps
\item Goldmania
\item Oreomystis
\item Rhodostethia
\item Falcipennis
\item Pachycoccyx
\item Cryptotympana
\item Tympanistalna
\item Cyrtoxipha
\item Afrixalus
\item Uperoleia
\item Urocitellus
\item Chalcorana
\item Aiolopus
\item Speothos

\end{enumerate}
\end{multicols}

\subsection{Held-out Species}
\begin{multicols}{3}
\begin{enumerate}
\setlength\itemsep{0.0em}
\tiny

\item Aethopyga shelleyi
\item Arachnothera dilutior
\item Sitta castanea
\item Carpodacus rodopeplus
\item Aethopyga ignicauda
\item Pachycephala soror
\item Herpsilochmus roraimae
\item Amazona dufresniana
\item Metallura aeneocauda
\item Thlypopsis fulviceps
\item Monarcha frater
\item Kleinothraupis reyi
\item Aplonis magna
\item Phylloscopus misoriensis
\item Agapornis pullarius
\item Amazona versicolor
\item Saltator cinctus
\item Xiphocolaptes falcirostris
\item Passer insularis
\item Chalcomitra balfouri
\item Arremonops tocuyensis
\item Atlapetes meridae
\item Colluricincla obscura
\item Saltator maxillosus
\item Philemon meyeri
\item Thamnophilus insignis
\item Aulacorhynchus whitelianus
\item Sirystes subcanescens
\item Sporophila nigrorufa
\item Zoothera mollissima
\item Thlypopsis inornata
\item Picumnus spilogaster
\item Columba arquatrix
\item Petrochelidon rufocollaris
\item Pyrrhura griseipectus
\item Myiothlypis chrysogaster
\item Thripophaga amacurensis
\item Herpsilochmus motacilloides
\item Progne dominicensis
\item Heliodoxa branickii
\item Asthenes arequipae
\item Gerygone fusca
\item Otus thilohoffmanni
\item Inezia subflava
\item Charadrius montanus
\item Petroica polymorpha
\item Symposiachrus vidua
\item Dicrurus lophorinus
\item Pycnonotus penicillatus
\item Melanerpes herminieri
\item Zosterops mysorensis
\item Oenanthe xanthoprymna
\item Artamus monachus
\item Caprimulgus pulchellus
\item Psarocolius cassini
\item Symposiachrus infelix
\item Zosterops cinereus
\item Circus cinereus
\item Geotrygon chrysia
\item Microspingus trifasciatus
\item Pternistis harwoodi
\item Ceblepyris caesius
\item Ficedula disposita
\item Treron affinis
\item Geokichla wardii
\item Campethera bennettii
\item Alcedo semitorquata
\item Buteo japonicus
\item Apus bradfieldi
\item Pterocles personatus
\item Melaniparus fringillinus
\item Poecile hypermelaenus
\item Circus buffoni
\item Pycnonotus blanfordi
\item Machlolophus aplonotus
\item Estrilda ochrogaster
\item Touit batavicus
\item Mirafra gilletti
\item Pternistis icterorhynchus
\item Accipiter collaris
\item Knipolegus lophotes
\item Nothoprocta taczanowskii
\item Pachycephala modesta
\item Vanellus tricolor
\item Caprimulgus andamanicus
\item Ardenna grisea
\item Mixornis kelleyi
\item Cinnyris johannae
\item Recurvirostra novaehollandiae
\item Sitta leucopsis
\item Petroica pusilla
\item Amazilia luciae
\item Melaniparus fasciiventer
\item Egretta picata
\item Columba pollenii
\item Rallus madagascariensis
\item Heliodoxa gularis
\item Carpodacus roseus
\item Zosterops chloronothos
\item Pachycephala lorentzi
\item Saucerottia cyanura
\item Cinclosoma marginatum
\item Bucco noanamae
\item Certhia nipalensis
\item Pachycephala lanioides
\item Carpodacus trifasciatus
\item Chorthippus acroleucus
\item Chlidonias albostriatus
\item Hirundo domicola
\item Falco concolor
\item Dryocopus schulzii
\item Rhyticeros undulatus
\item Quiscalus nicaraguensis
\item Cisticola brunnescens
\item Knipolegus cyanirostris
\item Ardenna carneipes
\item Lybius rubrifacies
\item Climacteris melanurus
\item Puffinus opisthomelas
\item Manorina melanotis
\item Celebesica abbotti
\item Otus mayottensis
\item Trachyphonus margaritatus
\item Oenanthe dubia
\item Chloropsis flavipennis
\item Ploceus alienus
\item Phalacrocorax varius
\item Ploceus pelzelni
\item Merops mentalis
\item Passer gongonensis
\item Myzomela cineracea
\item Pachycephala feminina
\item Brachypteryx sinensis
\item Lonchura flaviprymna
\item Ninox natalis
\item Myrmelastes caurensis
\item Buteo trizonatus
\item Apalis chariessa
\item Ficedula nigrorufa
\item Pica mauritanica
\item Anthreptes reichenowi
\item Sholicola major
\item Vireo osburni
\item Anas capensis
\item Ducula luctuosa
\item Lanius newtoni
\item Odontophorus dialeucos
\item Bostrychia olivacea
\item Cinnyris tsavoensis
\item Ploceus heuglini
\item Myzomela nigrita
\item Falco cherrug
\item Ixobrychus sturmii
\item Rhipidura semirubra
\item Haematopus chathamensis
\item Anthus brachyurus
\item Oenanthe lugens
\item Columba rupestris
\item Rhyticeros subruficollis
\item Zosterops vellalavella
\item Anthus sokokensis
\item Phaethornis idaliae
\item Picus dedemi
\item Muscicapa segregata
\item Cyanomitra bannermani
\item Polioptila facilis
\item Platysteira albifrons
\item Dicaeum pygmaeum
\item Puffinus assimilis
\item Rhipidura kubaryi
\item Ploceus katangae
\item Canis lupaster
\item Hyla andersonii
\item Ranoidea nudidigita
\item Ranoidea aurea
\item Litoria tyleri
\item Dendropsophus joannae
\item Okanagana occidentalis
\item Litoria latopalmata
\item Magicicada tredecassini
\item Orchelimum silvaticum
\item Oecanthus celerinictus
\item Empidonomus aurantioatrocristatus
\item Bufotes boulengeri
\item Oecanthus nigricornis
\item Myrmothera fulviventris
\item Psaltoda adonis
\item Rana dalmatina
\item Dendropsophus sanborni
\item Hyperolius stictus
\item Hyperolius pictus
\item Hyla eximia
\item Leptodactylus natalensis
\item Oecanthus californicus
\item Hyperolius parallelus
\item Gryllus cohni
\item Physeter macrocephalus
\item Eleutherodactylus unicolor
\item Gryllus bermudensis
\item Anas penelope

\end{enumerate}
\end{multicols}

\section{Evaluation on birdset}
We evaluate Top-1 accuracy on the datasets from the BirdSet benchmark. To match other models evaluated on BirdSet, which are constrained to predict one of the allowed labels, we use loss-based classification across all datasets and make predictions using scientific names. Our model achieves the highest average Top-1 accuracy, slightly surpassing Perch, demonstrating strong generalization from primarily focal recordings to soundscape recordings, and state-of-the-art performance for retrieval and classification on real-world bird datasets. 
\begin{table*}[ht]

\centering
{

{

\begin{tabular}{lcccccccc}
\toprule
 & POW & PER & NES & UHH & HSN & NBP & SNE & AVG \\ 
\midrule
EffNet   & 0.80 & 0.38 & \underline{0.49} & 0.42 & \textbf{0.59} & 0.63  & 0.67 & 0.57 \\
ConvNext & 0.75 & 0.36 & 0.45 & 0.44 & 0.52 & 0.64 & 0.65 & 0.54\\
AST      & 0.79 & 0.40 & 0.48 & 0.39 & 0.48 & 0.61 & 0.57 & 0.53 \\
EAT      & 0.69 & 0.32 & 0.46 & 0.40 & 0.47 & 0.61 & 0.58 & 0.50 \\
W2V2     & 0.72 & 0.34 & 0.47 & 0.51 & 0.50 & 0.65& 0.51 & 0.53\\
Perch    & \underline{0.85} & \underline{0.48} & \textbf{0.66} & \underline{0.57} & \underline{0.58} & \textbf{0.69} & \underline{0.69} & \underline{0.65} \\
\method & \textbf{0.95} & \textbf{0.62} & 0.47 & \textbf{0.60} & \underline{0.58} & \underline{0.66}  & \textbf{0.76} & \textbf{0.66} \\
\bottomrule
\end{tabular}
}
\small
\caption{Top-1 Accuracy results for each method on the datasets of BirdSet. Refer to the original paper~\citep{rauch2025birdset} for the details of compared baseline models.}
\label{tab:t1-acc-only}

}
\end{table*}

\section{Species Classification with Additional Context}\label{app:species_context}
We evaluate whether \method can improve species classification performance by incorporating additional context as text. The CBI dataset \citep{cornell2020}, derived from Xeno-canto, often contains metadata such as location and free-text notes written by recordists. We evaluate the model under three conditions: using audio alone, adding metadata (latitude, longitude, altitude when available, and geographic region), and further incorporating free-text notes. The model achieves an accuracy of 0.776 with audio alone, 0.792 with additional metadata, and 0.798 with both metadata and free-text notes, demonstrating that providing additional textual context can improve audio classification performance.

\section{Call Types and Transfer}

\begin{table}[H]
\begin{center}
{
\begin{tabular}{lrrrrrr}
\toprule
         Configuration & call-song &   multi &  call-song-unseen &  multi-unseen \\
\midrule
    \method & \textbf{0.871}  & \textbf{0.667} & \textbf{0.769} & \textbf{0.678} \\
BioLingual & 0.658  & 0.303 & 0.665 & 0.419 \\
\bottomrule
\end{tabular}
}
\end{center}
\caption{Accuracy of call vs. song classification (call-song), multi-call classification (multi), and the generalization of these tasks to unseen taxa (call-song-unseen, multi-unseen.)}
\label{tab:call-type-transfer}
\end{table}

We further evaluate the classification of bird call types and the transfer of this task across species. We test the model on call vs. song prediction as well as call-type prediction for multiple classes (call, song, flight call, alarm call, begging call, and drumming). We then test if these tasks can be transferred to unseen taxa. The call-song-unseen and multi-unseen datasets evaluate the same tasks described above, but evaluated on the held-out taxa used to test unseen species, unseen genus, and unseen family. In addition to achieving state-of-the-art results on these tasks, the results transfer strongly to unseen taxa, outperforming BioLingual---even when these taxa were held out from \method but not from BioLingual.

\section{Ablation on Unfreezing BEATs}
\begin{table}[H]
\begin{center}
{
\begin{tabular}{lrrrrrr}
\toprule
         Configuration & watkins &   cbi &  unseen-species &  unseen-genus & unseen-family\\
\midrule
    BEATs-unfrozen & \textbf{0.723} & \textbf{0.680} & \textbf{0.320} & \textbf{0.124}  & \textbf{0.306} \\
BEATs-frozen & {0.490} & {0.401} & {0.184} & {0.073}  & {0.186} \\
\bottomrule
\end{tabular}
}
\end{center}
\caption{Zero-shot classification results with BEATs unfrozen vs. frozen. Both models are trained on stage-1 tasks for 150k steps. We report accuracy on species classification tasks, with unseen taxa tasks predicted using taxonomic names.}
\label{tab:freeze-beats}
\end{table}

\section{Speech+music ablation: full results}
\begin{table}[H]
\begin{center}
{
\begin{tabular}{lrrrrrrrrr}
\toprule
         Model & esc50 &  watkins &   cbi &  humbugdb &  dcase &  enabirds &  hiceas &   rfcx &   gibbons  \\
\midrule
    base & {0.570} & {0.788} & {0.748} & {0.093} &  {0.107} & { 0.299} &   {0.415} & {0.038} & {0.011} \\
base w/o speech or music & {0.605} & {0.773} & {0.750} & {0.152} &  {0.040} & { 0.293} &   {0.417} & {0.038} & {0.012} \\
\bottomrule
\end{tabular}
}
\end{center}
\caption{Zero-shot classification and detection results on \newbeans. Base model was trained on all stage-2 training tasks, while ``base w/o speech or music'' is an ablation removing both speech and music tasks from training data. Both models were trained for 150k steps. We used accuracy for classification, and F1 for detection tasks.}
\label{tab:speech-beans}
\end{table}

\begin{table}[H]
\begin{center}
{\small
\begin{tabular}{lrrrrrrrrrr}
\toprule
         Model &  unseen-species   & unseen-genus & unseen-family & lifestage   & call-type &  captioning  & zf-indv \\
\midrule
base & {0.322} & {0.139} & {0.239} & {0.702} & {0.863} & {0.501} & {0.677} \\
base w/o speech or music & {0.354} & {0.137} & {0.330} & {0.690} & {0.852} & {0.503} & {0.500} \\
\bottomrule
\end{tabular}
}
\end{center}
\caption{Zero-shot results on new tasks introduced in \newbeans. Base model was trained on all stage-2 training tasks, while base w/o speech or music is an ablation removing both speech and music tasks from training data. Both models were trained for 150k steps. We report accuracy for classification, and SPIDEr~\citep{sharif2018learning} for captioning.}
\label{tab:speech-added}
\end{table}

\newpage
\section{Training data}
\begin{figure*}[htb]
\begin{center}
\includegraphics[scale=0.095]{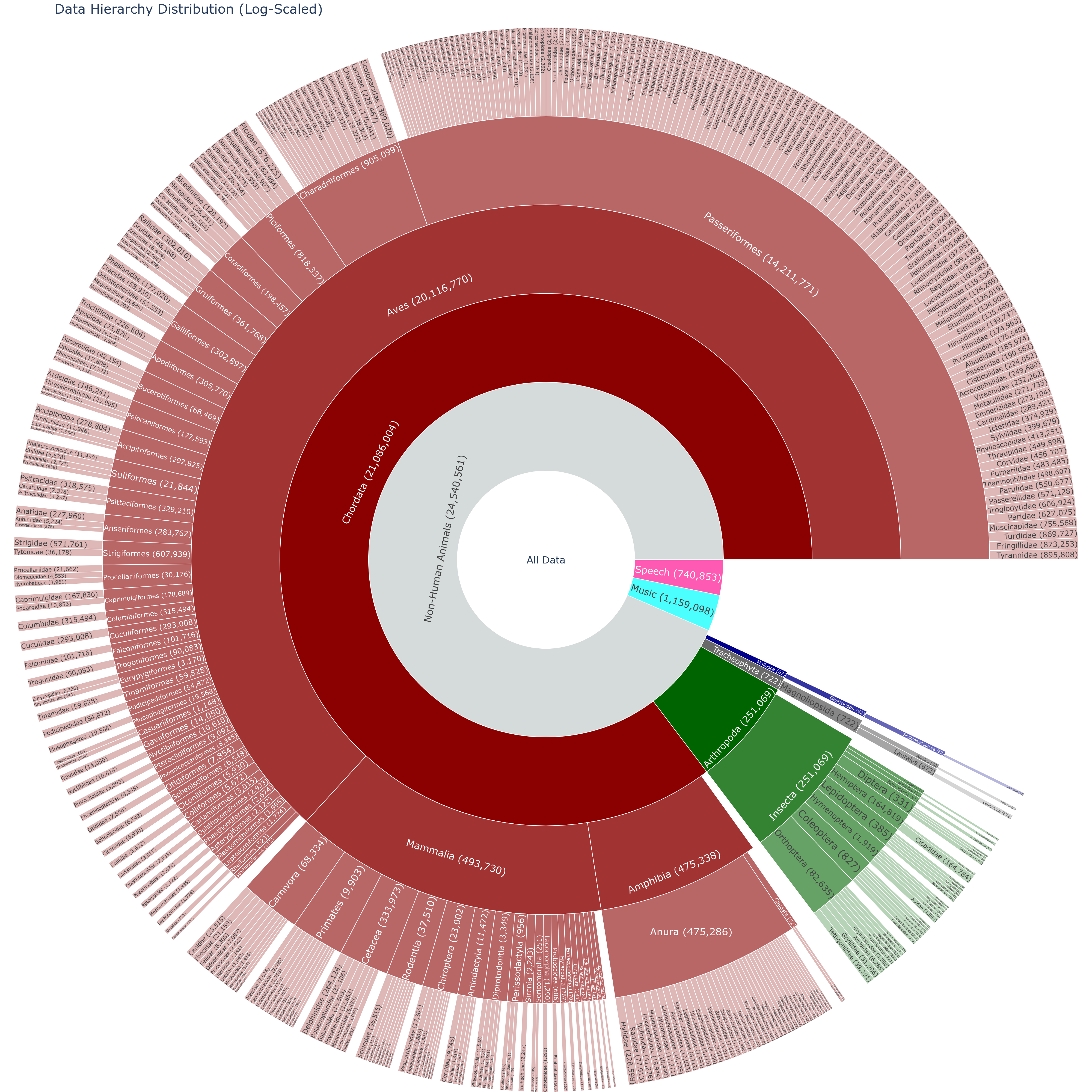} 
\caption{Data composition across training samples including the distribution for the main data types and phylum, class, and order for non-human animals. The counts represent prompts rather than audio files i.e. various prompts may be derived from the a single audio file.}
\label{fig:training_data}
\end{center}
\end{figure*}

\end{document}